%% file: D2D_MIMO_multicell_WSA.tex
\documentclass[conference]{IEEEtran}
\usepackage[algo2e]{algorithm2e}
\usepackage[super]{nth}
\usepackage{cite}
\usepackage{graphicx}  
\usepackage{amsmath}  
\usepackage{subfigure}
\usepackage{algorithm}
\usepackage[latin1]{inputenc}
\usepackage[T1]{fontenc}
\usepackage{times}
\usepackage{tabularx}
\usepackage{graphicx}
\usepackage{stfloats}
\usepackage{fancybox}
\usepackage{verbatim}
\usepackage{longtable}
\usepackage{url}
\usepackage{boxedminipage}
\usepackage{subfigure}
\usepackage{amssymb}
\usepackage{tabu}
\usepackage{pdfpages}
\usepackage{epstopdf}
\usepackage{cite}
\usepackage{algpseudocode}
\usepackage{optidef}

\input{commands}

\IEEEoverridecommandlockouts

\begin{document}
	
	\title{Power Control for D2D Underlay in \\ Multi-cell Massive MIMO Networks}
	
	\author{Amin Ghazanfari, Emil Bj{\"o}rnson, and Erik G. Larsson\\
		Department of Electrical Engineering (ISY), Link{\"o}ping University, Sweden\\
		Email:\{amin.ghazanfari, emil.bjornson, erik.g.larsson\}@liu.se
		\thanks{This paper was supported by the European Union's Horizon 2020 research
			and innovation programme under grant agreement No 641985 (5Gwireless).}}
	
	\maketitle
	\begin{abstract}
		This paper proposes a new power control and pilot allocation scheme for device-to-device (D2D) communication underlaying a multi-cell massive MIMO system.  In this scheme, the cellular users in each cell get orthogonal pilots which are reused with reuse factor one across cells, while the D2D pairs share another set of orthogonal pilots. We derive a closed-form capacity lower bound for the cellular users with different receive processing schemes. In addition, we derive a capacity lower bound for the D2D receivers and a closed-form approximation of it. 
		Then we provide a power control algorithm that maximizes the minimum spectral efficiency (SE) of the users in the network. Finally, we provide a numerical evaluation where we compare our proposed power control algorithm with the maximum transmit power case and the case of conventional multi-cell massive MIMO without D2D communication. 
		Based on the provided results, we conclude that our  proposed scheme increases the
		sum spectral efficiency of multi-cell massive MIMO networks.
	\end{abstract}
	
	\section{Introduction}
	D2D underlay communication and massive MIMO are two promising technologies that will appear in 5G networks \cite{WWB5G}. D2D underlay communication enhances the spectrum utilization by reusing the cellular resources for direct communication between D2D pairs when the transmitter and receiver are closely located. It provides benefits such as offloading gains for cellular networks, higher data rate and lower transmission power between D2D users due to short-range communication \cite{asadi2014survey,doppler2009device,lin2014overview}. Hence, D2D underlay communication increases spectral and energy efficiency of the cellular networks. These benefits come at the cost of causing extra interference to the cellular users (CUs) \cite{janis2009interference,AminICC,fodor2012design}. Massive MIMO improves spectral and energy efficiency of cellular networks by utilizing a large number of antennas at each base station (BS). It offers multiplexing gains and spatial interference suppression \cite{redbook,bjornson2017massive,larsson2014massive}. The latter seems to address the shortcoming of D2D underlay communication. Hence, the combination of these two technologies has received considerable attention. In prior works, D2D underlaying the uplink (UL) of massive MIMO systems has been investigated mainly for single-cell setups \cite{liu2017pilot, xu2016power, xu2017pilot}. In these papers, the transmitter and receiver of a D2D pair are always physically located in the same cell. This simplifies the resource allocation since each D2D pair shares resources with a specific cell. The multi-cell scenario has only been studied in \cite{lin2015interplay}. The authors in \cite{lin2015interplay} investigate the interplay between massive MIMO and underlaid D2D for uplink data transmission in a multi-cell massive MIMO setup where the number of D2D pairs in each cell follows a Poisson distribution. A detailed analysis is provided for the case of perfect channel state information (CSI), in which they study the asymptotic and non-asymptotic SE of CUs and D2D pairs. They also perform channel estimation based on an orthogonal pilot training scheme that allocates orthogonal pilots to CUs and limited number of D2D pairs nearest to the location of the BS in their cell. In this case, they study the asymptotic and non-asymptotic SE of cellular users only. It is assumed that all users transmit with a predefined transmit power.

	In contrast to most prior works, in this paper, we investigate a multi-cell massive MIMO setup with D2D underlaying  where the D2D transmitters and receivers have arbitrary locations. It is assumed that the CUs in a cell have orthogonal pilots that are reused with reuse factor one between cells. In addition, the system consists of D2D pairs that are not belonging to any cell and share a network-wide set of orthogonal pilots. We investigate the channel estimation at cellular BSs as well as D2D receivers and derive a closed-form capacity lower bound for the CUs with either maximum-ratio (MR) and zero-forcing (ZF) processing. In addition, we derive a capacity lower bound for D2D communication. This lower bound is not in closed form as it contains an expectation with respect to the small-scale fading. Hence, to provide a tractable power control scheme, we derive a closed-form approximation of the capacity lower bound. Furthermore, we formulate and solve a power control problem, which jointly optimizes the power of CUs and D2D pairs to guarantee max-min fairness performance for cellular and D2D communications. Finally, we provide a numerical performance evaluation of our proposed algorithm.
	
	\section{System Model}
	\label{SystemModel}
\begin{figure}[htpb]
		\centering
		\centerline{\includegraphics[width=8cm]{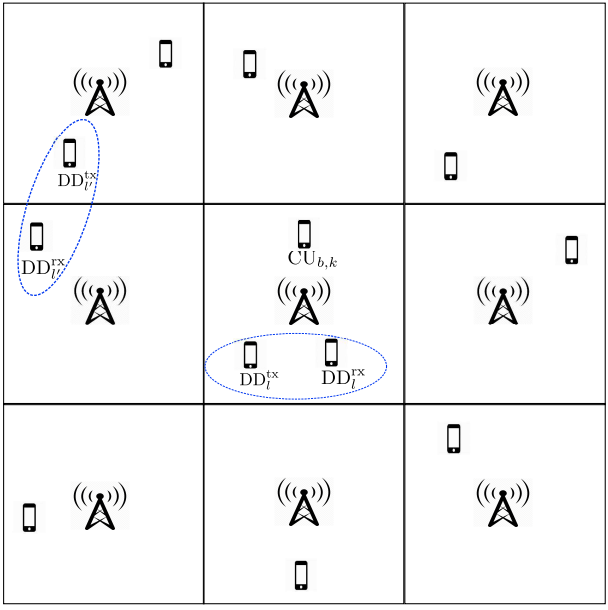}}
		\caption{Network setup.}
		\label{fig:model} \vspace{4mm}
	\end{figure}

	In this paper, we consider a multi-cell massive MIMO system consisting of $B$ BSs, each equipped with an array of $M$ antennas and each BS serves $K$ single-antenna CUs. The system also contains $L$ D2D pairs that are spread over the whole network and do not belong to any specific cell. The network model is illustrated in Fig.~\ref{fig:model}. In this setup, we investigate D2D communication underlaying the UL transmission of the multi-cell massive MIMO system. The wireless channels are varying over time and frequency, which we model by block fading. We define the coherence interval of a channel as the time-frequency block in which the channel is constant and flat-fading and its size is $\tau_c$ symbols \cite{redbook,bjornson2017massive}. The channels change independently from one block to another according to a stationary ergodic random process. The size of the coherence interval depends on two factors, which are the coherence time $T_{c}$ and the coherence bandwidth $B_{c}$, and $\tau_c = T_{c} B_{c}$.
	In the proposed setup,  ${\vec h}^{b,\rm c}_{b',k}\sim \CN({\vec{0}},\beta^{b,\rm c}_{b',k} {\vec I}_{M})$ is the channel response between the BS $b$ and CU $k$ in cell $b'$ and ${\vec h}^{b,\rm d}_{l} \sim \CN({\vec{0}},\beta^{b,\rm d}_l {\vec I}_M)$ is the channel response between D2D transmitter~$l$~and BS $b$. In addition, ${g}^{l,\rm c}_{b,k} \sim \CN(0,\beta^{l,\rm c}_{b,k}) $ denotes the channel response between CU~$k$~located in cell $b$ and D2D receiver~$l$, while ${g}^{l,\rm d}_{l'} \sim \CN(0,\beta^{l,\rm d}_{l'})$ denotes the channel between D2D transmitter~$l'$~and D2D receiver~$l$.
	
	\vspace{4mm}

	In our network setup, it is assumed that the BSs and D2D receivers have no prior channel state information at the beginning of a coherence interval; hence, channel estimation is carried out in each coherence interval. The communication therefore consists of two phases: UL pilot transmission and UL data transmission. To enable channel estimation, the CUs and D2D transmitters transmit pilot sequences of length $\tau$ in the pilot transmission phase and the remaining $\tau_{c} - \tau$ symbols are utilized in the data transmission phase. In this setup, we can construct $\tau$ orthogonal pilot sequences that are vectors of size $\tau$. It is assumed that we have $K$ orthogonal pilots for CUs that are reused in each cell and we have another set of $N$ orthogonal pilots designated for the $L\geq N$ D2D pairs. For the aforementioned setup we have $\tau = N+K$.
	
	\subsection{Uplink data transmission}
	The received signal at BS $b$ during the data transmission  is
	\begin{equation}
	\begin{aligned}\label{eq:data BS-cu}
	{\vec y}^{\rm c}_{b}&= \sum\limits_{k=1}^{K}\sqrt{p^{\rm c}_{b,k}} {\vec h}^{b,\rm c}_{b,k} s^{\rm c}_{b,k}+\sum\limits_{b'=1,b'\neq b}^{B}\sum\limits_{k=1}^{K}\sqrt{p^{\rm c}_{b',k}} {\vec h}^{b,\rm c}_{b',k} s^{\rm c}_{b',k}\\
	&+\sum\limits_{l=1}^{L}\sqrt{{p^{\rm d}_{l}}}{\vec h}^{b,\rm d}_{l} s^{\rm d}_{l}+{\vec w}_{b}\\
	\end{aligned}
	\end{equation}
	and the received signal at D2D receiver~$l$~is 
	\begin{equation}
	\begin{aligned}\label{eq:data BS-dd}
		{y}^{\rm d}_l&= \sum\limits_{b'=1}^{B}\sum\limits_{k=1}^{K}\sqrt{p^{\rm c}_{k}} {g}^{l,\rm c}_{b,k}s^{\rm c}_{b,k} +\sum\limits_{l'=1}^{L}\sqrt{{p^{\rm d}_{l'}}}{g}^{l,\rm d}_{l'}s^{\rm d}_{l'}+{w}_l,\\
	\end{aligned}
	\end{equation}
	where~$p^{\rm c}_{b,k}$~and~$p^{\rm d}_{l'}$~are the transmit powers for data transmission of the CU~$k$ in cell $b$ and the D2D transmitter $l'$, respectively. ${\vec w}_{b}\sim\CN({\vec{0}},{\vec{I}}_M)$~and~${w}_l\sim\CN(0,1)$~indicate the normalized additive white Gaussian noise at BS $b$ and the $l$th D2D receiver, respectively. Also,~$s^{\rm c}_{b,k}$~and~$s^{\rm d}_{l'}$~denote zero mean and unit variance data symbols transmitted from CU~$k$ in cell $b$ and D2D transmitter $l'$, respectively.
	\section{Analysis of spectral efficiency}
	In this section, we derive the SEs achieved when using the communication setup defined in Section~\ref{SystemModel}. Pilot contamination arises between CUs due to the pilot reuse between cells and pilot contamination also arises between the D2D pairs that are using the same pilot.
	
	\subsection{Pilot transmission and channel estimation}
	We denote the matrix of pilot sequences as $\mathbf{\boldsymbol{\Phi}} = [\mathbf{\boldsymbol{\Phi}}^{\rm c}~\mathbf{\boldsymbol{\Phi}}^{\rm d}]$ and it has size $\tau \times (K+N)$. The matrices $\mathbf{\boldsymbol{\Phi}}^{\rm c} = [\mathbf{\boldsymbol{\phi}}^{\rm c}_{1},\ldots,\mathbf{\boldsymbol{\phi}}^{\rm c}_{K}] \in \mathbb{C}^{\tau \times K}$ and $\mathbf{\boldsymbol{\Phi}}^{\rm d} = [\mathbf{\boldsymbol{\phi}}^{\rm d}_{1},\ldots,\mathbf{\boldsymbol{\phi}}^{\rm d}_{N}] \in \mathbb{C}^{\tau \times N}$ contain the orthogonal unit-norm pilot sequences assigned for CUs and D2D pairs, respectively. The pilot $\mathbf{\boldsymbol{\phi}}^{\rm c}_{k}$ is used by CU~$k$ in each of the cells.
	The sets $n_1,\ldots,n_N$  contains the indicies of D2D pairs that are using pilots $\mathbf{\boldsymbol{\phi}}^{\rm d}_1,\ldots,\mathbf{\boldsymbol{\phi}}^{\rm d}_N$, respectively.

	Each BS multiplies the received signal matrix in the pilot transmission phase with each of the pilot signals to despread the signals. Hence, the received pilot signals at BS $b$ from its CU $k$ after despreading is~\cite{redbook,bjornson2017massive}
	\begin{equation}
	\begin{aligned}\label{eq:pilot BS-cu}
		{\vec y}^{b,\rm c}_{b,k}&= \sum\limits_{b'=1}^{B}\sqrt{\tau p^{\rm p,c}_{b',k}} {\vec h}^{b,\rm c}_{b',k}+{\vec w}^{b,\rm c}_{k},\quad k=1,\ldots,K,\\ 
	\end{aligned}
	\end{equation}
	where~$p^{\rm p,c}_{b,k}$~denotes the pilot transmit power used by CU~$k$~in cell $b$ and~${\vec w}^{b,\rm c}_k\sim\CN({\vec{0}},{\vec{I}}_{M})$ is the normalized additive white Gaussian noise at the BS $b$. The received pilot signal at BS $b$ from all D2D transmitters is~\cite{redbook,bjornson2017massive}
	\begin{equation}
	\begin{aligned}
		{\vec Y}^{b,\rm d}&= \sum\limits_{i=1}^{N}\sum_{l \in n_i}\sqrt{\tau {p^{p,\rm d}_{l}}}{\vec h}^{b,\rm d}_{l} (\boldsymbol{\phi}^{{\rm d}}_i)^H+{\vec W}^{b,\rm d}, \\
	\end{aligned}
	\end{equation}
	and despreading by multiplication with $\mathbf{\boldsymbol{\Phi}}^{\rm d}$ yields
	\begin{equation}
	\begin{aligned}\label{eq:pilot BS-DD}
		{\vec Y}^{b,\rm d} \boldsymbol{\Phi}^{\rm d}&= \sum\limits_{i=1}^{N}\sum_{l\in n_i}\sqrt{\tau {p^{p,\rm d}_{l}}}{\vec h}^{b,\rm d}_{l}(\boldsymbol{\phi}^{{\rm d}}_i)^H\boldsymbol{\Phi}^{{\rm d}}+{\vec W'}^{b,\rm d}, \\
	\end{aligned}
	\end{equation}
	where $p^{\rm p,d}_{l}$~is the transmit power of D2D transmitter~$l$~for pilot transmission and~${\vec W'}^{b,\rm d} = {\vec W}^{b,\rm d}{\Phi}^{{\rm d}} \in \mathbb{C}^{N\times \tau}$~is the normalized additive white Gaussian noise at the BS $b$ which has independent entries having the distribution $\CN(0,1)$.  
	The MMSE estimates of the channel vectors from CU~$k$ at BS $b$ is \cite{kailath2000linear,kay1993fundamentals}
	\begin{equation}
	\begin{aligned}\label{eq:mmse BS-cu}
		\hat{{\vec h}}^{b,\rm c}_{b,k} =  \frac{\sqrt{\tau p^{\rm p,c}_{b,k}} \beta^{b,\rm c}_{b,k}}{1+ \tau \sum\limits_{b'=1}^{B} p^{\rm p,c}_{b',k} \beta^{b,\rm c}_{b',k}}\textbf{y}^{b,\rm c}_{b,k}.\\
	\end{aligned}
	\end{equation}
	The MMSE estimate at BS $b$ of the sum of the channel vectors $\sum_{l\in n_i} \sqrt{\smash[b]{\tau p^{\rm p, d}_{l}}} {\vec h}^{b,\rm d}_{l}$ from D2D transmitters in set $n_i$  is 
	\begin{equation}
	\begin{aligned} \label{eq:mmse BS-dd-ni}
		\hat{{\vec h}}^{b,n_i}_{} = \frac{\sum\limits_{l' \in n_i} {\tau p^{\rm p, d}_{l'}} \beta^{b,\rm d}_{l'}}{1+ \sum\limits_{l'\in n_i}\tau p^{p,\rm d}_{l'} \beta^{b,\rm d}_{l'}} \left[{\vec Y}^{b,\rm d} \mathbf{\boldsymbol{\phi}}^{d}_{i}\right].\\
	\end{aligned}
	\end{equation}
	In addition, the MMSE estimates of the channel vectors from  D2D transmitter~$l \in n_i$ at BS $b$ is
	\begin{equation}
	\begin{aligned}\label{eq:mmse BS-dd}
		\hat{{\vec h}}^{b,\rm d}_{l} = \frac{\sqrt{\tau p^{\rm p, d}_{l}} \beta^{b,\rm d}_{l}}{1+ \sum\limits_{l'\in n_i}\tau p^{p,\rm d}_{l'} \beta^{b,\rm d}_{l'}} \left[{\vec Y}^{b,\rm d} \mathbf{\boldsymbol{\phi}}^{\rm d}_i\right].\\
	\end{aligned}
	\end{equation}
	The mean square of the channel estimates at the BS are $\mathbb{E}[\|\hat{{\vec h}}^{b,\rm c}_{b,k}\|^2]= \gamma^{b,\rm c}_{b,k} M$, $\mathbb{E}[\|	\hat{{\vec h}}^{b,n_i}_{}\|^2]= \gamma^{b,n_i}_{} M$, and $\mathbb{E}[\|	\hat{{\vec h}}^{b,\rm d}_{l}\|^2]=\gamma^{b,\rm d}_{l} M$, 
	in which we have used
	\begin{equation}
		\gamma^{b,\rm c}_{b,k} = \frac{\tau p^{\rm p,c}_{b,k} \left({\beta^{b,\rm c}_{b,k}}\right)^2}{1+\tau \sum\limits_{b'=1}^{B} p^{\rm p,c}_{b',k} \beta^{b,\rm c}_{b',k}},
	\end{equation}
	\begin{equation}
	\gamma^{b,n_i}_{} = \frac{\tau \left(\sum\limits_{l' \in n_i} \sqrt{p^{p,\rm d}_{l'}} {\beta^{b,\rm d}_{l'}}\right)^2}{1+\sum\limits_{l'\in n_i}\tau p^{\rm p, d}_{l'} \beta^{b,\rm d}_{l'}},
	\end{equation}
	\begin{equation}
		\gamma^{b,\rm d}_{l} = \frac{\tau p^{p,\rm d}_{l} \left({\beta^{b,\rm d}_{l}}\right)^2}{1+\sum\limits_{l'\in n_i}\tau p^{\rm p, d}_{l'} \beta^{b,\rm d}_{l'}}.
	\end{equation}
	Note that the channel estimate $\hat{{\vec h}}^{b,\rm d}_{l} $ of  D2D transmitter~$l \in n_i$ is a scaled version of $\hat{{\vec h}}^{b,n_i}_{}$, which is a property that we will utilize later. 
	
	The D2D receivers apply the same procedure for channel estimation. Therefore, the received signals at D2D receiver~$l$~from the pilot transmission of CU~$k$ located in cell $b$ and D2D transmitter~$l' \in n_i$~after despreading the signals are
	\begin{equation}
	\begin{aligned}\label{eq:pilot DD-cu}
		{y}^{l,\rm c}_{b,k}&=  \sum\limits_{b'=1}^{B}\sqrt{\tau p^{l,\rm p,c}_{b',k}} {g}^{l,\rm c}_{b',k}+{w}^{\rm c}_{l}, \\
	\end{aligned}
	\end{equation}
	\begin{equation}
	\begin{aligned}\label{eq:pilot dd}
		{y}^{l,\rm d}_{l'}&= \sum\limits_{l''\in n_{i}}\sqrt{\tau {p^{\rm p,d}_{ l''}}}{g}^{l,\rm d}_{l''} +{ w}^{\rm d}_l, \\
	\end{aligned}
	\end{equation}
	where ${w}^{\rm d}_l\sim\CN(0,1) $ and ${ w}^{\rm c}_l \sim\CN(0,1)$ are the normalized additive white Gaussian noise terms at D2D receiver~$l$. 
	The pilot transmission of all transmitters, i.e., cellular and D2D, are used for channel estimation and the MMSE estimates of the channels from CU~$k$~located in cell $b$ and from D2D transmitter~$l' \in n_i$~are respectively given by
	\begin{equation}
	\begin{aligned}\label{eq:mmse dd-cu}
		\hat{g}^{l,\rm c}_{b,k} = \frac{\sqrt{\tau p^{l,\rm p,c}_{b,k}} \beta^{l,\rm c}_{b,k}}{1+ \sum\limits_{b'=1}^{B}\tau p^{l,\rm p,c}_{b',k} \beta^{l,\rm c}_{b',k}} {y}^{l,\rm c}_{b,k}, \\
	\end{aligned}
	\end{equation}
	\begin{equation}
	\begin{aligned}\label{eq:mmse dd}
		\hat{g}^{l,\rm d}_{l'} = \frac{\sqrt{\tau p^{\rm p,d}_{l'}} \beta^{l,\rm d}_{l'}}{1+ \sum\limits_{l''\in n_i}\tau p^{\rm p,d}_{l''} \beta^{l,\rm d}_{l''}} {y}^{l,\rm d}_{l'}.\\
	\end{aligned}
	\end{equation}
	In addition, the mean square of the channel estimates at D2D receiver~$l$~are
	\begin{equation}
	\begin{aligned}
		\mathbb{E}\left[|\hat{g}^{l,\rm c}_{b,k}|^2\right] = {\gamma^{l,\rm c}_{b,k}}=  \frac{\tau p^{l,\rm p,c}_{b,k} \left({\beta^{l,\rm c}_{b,k}}\right)^2}{1+ \sum\limits_{b'=1}^{B}\tau p^{l,\rm p,c}_{b',k} \beta^{l,\rm c}_{b',k}},
	\end{aligned}
	\end{equation}
	\begin{equation}
	\begin{aligned}
		\mathbb{E}\left[|\hat{g}{^{l,\rm d}_{l'}}|^2\right] = \gamma^{l,\rm d}_{l'}=  \frac{\tau p^{\rm p,d}_{l'} \left({\beta^{l,\rm d}_{l'}}\right)^2}{1+ \sum\limits_{l''\in n_i}\tau p^{\rm p,d}_{l''} \beta^{l,\rm d}_{l''}}.
	\end{aligned}
	\end{equation}
	\subsection{Spectral efficiency with MR processing}
	In this subsection, we assume that MR processing is applied at all BSs to detect the signals of their own users. We utilize the use-and-then-forget technique~\cite[Ch.3]{redbook} to lower bound the capacity of each of the CUs, using the capacity bound for a deterministic channel with additive white non-Gaussian noise provided in~\cite[Sec.~2.3]{redbook}. We get the following lower bound on the capacity of CU~$k$ in cell $b$:
	\begin{equation}
	\begin{aligned}\label{eq:MRC-bound}
		R^{b,\rm c}_{b,k}= &  
		\left(1 - \frac{\tau }{\tau_c}\right)\log_2 \left(1+\frac{M p^{\rm c}_{b,k} \gamma^{b,\rm c}_{b,k}}{ I^{\rm mr}_{b,k} }\right),\\
	\end{aligned}
	\end{equation}
	where
	\begin{equation}
	\begin{aligned}
		I^{\rm mr}_{b,k} &=  1+ \sum\limits_{b'=1}^{B}\sum\limits_{k'=1}^{K} p^{\rm c}_{b',k'} \beta^{b,\rm c}_{b',k'} + \sum\limits_{l=1}^{L}p^{\rm d}_{l}  \beta^{b,\rm d}_l \\
		&+M\sum\limits_{b'=1,b'\neq b}^{B} p^{\rm c}_{b',k} \gamma^{b,\rm c}_{b',k}.
	\end{aligned}
	\end{equation}
	where $\frac{M p^{\rm c}_{b,k} \gamma^{b,\rm c}_{b,k}}{ I^{\rm mr}_{b,k}}$ is the effective SINR expression in which the numerator indicate the beamforming gain of the desired signal at BS $b$. In $	I^{\rm mr}_{b,k}$, the first term is the noise variance, the second term is non-coherent interference form CUs in contaminating cells, the third term is non-coherent interference from D2D communication, and the last term is coherent interference from CUs in other cell that are sharing pilot sequence with user $k$.
	\subsection{Spectral efficiency with ZF processing}
	In this subsection, we consider ZF processing at all BSs, instead of MR processing, to cancel interference between the users. In addition to cancelling intracell interference, we also cancel interference from the D2D transmitters. For ease of notation, we rewrite~\eqref{eq:data BS-cu}~in matrix form as
	\begin{equation}
	\begin{aligned}\label{eq:data BS-cu-matrix}
		{\vec y}^{\rm c}_{b} =& {\vec{H}}^{b,\rm c}_{b} {\vec{D}}^{1/2}_{{\vec{p}}^{\rm c}_{b}}{\vec{s}}^{\rm c}_{b}+\sum\limits_{b'=1,b'\neq b}^{B} {\vec{H}}^{b,\rm c}_{b'} {\vec{D}}^{1/2}_{{\vec{p}}^{\rm c}_{b'}}{\vec{s}}^{\rm c}_{b'}
		\\& +{\vec{H}}^{b,\rm d} {\vec{D}}^{1/2}_{\vec{p}^{\rm d}} {\vec{s}}^{b,\rm d}
		{\vec{w}}_{b},
	\end{aligned}
	\end{equation}
	where~${\vec{H}}^{b,\rm c}_{b'} = [{\vec h}^{b,\rm c}_{b',1},\ldots,{\vec h}^{b',\rm c}_{b,K}]$ is the channel matrix of the $K$ CUs located in cell $b'$ to BS $b$ and~${\vec{H}}^{b',\rm d} = [{\vec h}^{b',\rm d}_{1},\ldots,{\vec h}^{b',\rm d}_{L}]$ is the channel matrix of the D2D transmitters to BS $b'$. ${\vec{D}}_{{\vec{p}}^{\rm c}_{b'}}$~and~${\vec{D}}_{{\vec p}^{\rm d}}$~are diagonal matrices indicating the transmit power of CUs in cell $b'$ and D2D transmitters, respectively, where ${\vec{p}}^{\rm c}_{b'} = \left[ p^{\rm c}_{b'1},\ldots,p^{\rm c}_{b'K}\right]^{T}$ and ${\vec{p}}^{\rm d} = \left[ p^{\rm d}_{1},\ldots,p^{\rm d}_{L}\right]^{T}$ are the vectors on the diagonals. The channel matrices in~\eqref{eq:data BS-cu-matrix}~can be replaced with the corresponding estimated channel matrices $\hat{\vec{H}}^{b,\rm c}_{b'} = [\hat{{\vec h}}^{b,\rm c}_{b'1},\ldots,\hat{{\vec h}}^{b,\rm c}_{b',K}]$~and~$\hat{\vec{H}}^{b,\rm d} = [\hat{{\vec h}}^{\rm d}_1,\ldots,\hat{{\vec h}}^{\rm d}_{L}]$, and estimation error matrices defined as $\tilde{\vec{H}}^{b,\rm c}_{b'}= \hat{\vec{H}}^{b,\rm c}_{b'} - {\vec{H}}^{b,\rm c}_{b'} = [\tilde{{\vec h}}^{b,\rm c}_{b'1},\ldots,\tilde{{\vec h}}^{b,\rm c}_{b',K}]$~and~$\tilde{\vec{H}}^{b,\rm d} = \hat{\vec{H}}^{b,\rm d} - {\vec{H}}^{b,\rm d}= [\tilde{{\vec h}}^{b,\rm d}_1,\ldots,\tilde{{\vec h}}^{b,\rm d}_{L}]$. Hence, \eqref{eq:data BS-cu-matrix} is rewritten as 
	\begin{equation}
	\begin{aligned}\label{eq:data BS-cu matrix}
		&{\vec y}^{\rm c}_{b}= \hat{\vec{H}}^{b,\rm c}_{b} {\vec{D}}^{1/2}_{{\vec{p}}^{\rm c}_{b}}{\vec{s}}^{\rm c}_{b}+\sum\limits_{b'=1,b'\neq b}^{B} \hat{\vec{H}}^{b,\rm c}_{b'} {\vec{D}}^{1/2}_{{\vec{p}}^{\rm c}_{b'}}{\vec{s}}^{\rm c}_{b'} + \hat{\vec{H}}^{b,\rm d} {\vec{D}}^{1/2}_{\vec{p}^{\rm d}} {\vec{s}}^{b,\rm d}\\
		&+{\vec{w}}_{b} -\tilde{\vec{H}}^{b,\rm c}_{b} {\vec{D}}^{1/2}_{{\vec{p}}^{\rm c}_{b}}{\vec{s}}^{\rm c}_{b} -\!\!\!\sum\limits_{b'=1,b'\neq b}^{B} \!\!\!\tilde{\vec{H}}^{b,\rm c}_{b'} {\vec{D}}^{1/2}_{{\vec{p}}^{\rm c}_{b'}}{\vec{s}}^{\rm c}_{b'}-\tilde{\vec{H}}^{b,\rm d} {\vec{D}}^{1/2}_{\vec{p}^{\rm d}} {\vec{s}}^{b,\rm d}.
	\end{aligned}
	\end{equation}
	As mentioned previously, the channel estimates at BS $b$ for D2D transmitters that use the same pilot are parallel vectors. Hence, the collection of $N$ vector $\hat{\vec{h}}^{b,n_1},\ldots,\hat{\vec{h}}^{b,n_N}$ from \eqref{eq:mmse BS-dd-ni} contains scaled version of all the $L$ channel estimates $\hat{{\vec h}}^{b,\rm d}_{1}, \ldots,\hat{{\vec h}}^{b,\rm d}_{L}$.
	Therefore, we construct the $M \times (K+N)$ matrix 
	\begin{equation}
	\begin{aligned}
		\hat{\vec{H}}^{b}_{b} &= [\hat{\vec{H}}^{b,\rm c}_{b}\quad\hat{\vec{h}}^{b,n_1}\,\,\ldots\,\,\hat{\vec{h}}^{b,n_N}]
	\end{aligned}
	\end{equation}
	to describe the channel directions that are relevant for interference mitigation at BS $b$.
	We can write this matrix as 
	\begin{equation}
	\begin{aligned}
		\hat{\vec{H}}^{b}_{b} &= {\vec{Z}} {\vec{D}}^{1/2}_{\bgamma^{b}},
	\end{aligned}
	\end{equation}
	where the matrix ${\vec Z}$ has i.i.d.~$\CN(0,1)$ elements and ${\vec{D}}_{\bgamma^{b}}$ is a diagonal matrix with ${{\bgamma}^{b}} = [\gamma^{b,\rm c}_{b,1},\ldots,\gamma^{b,\rm c}_{b,K},\gamma^{b,n_1},\ldots,\gamma^{b,n_N}]$ at the diagonal.
	By using this matrix, we define the ZF detection matrix as 
	\begin{equation}
	\begin{aligned}
		{\vec V}^{b}&= \hat{\vec{H}}^{b}_{b} \left(\left({\hat{\vec{H}}_{b}}^{b}\right)^{\rm H} \hat{\vec{H}}^{b}_{b} \right)^{-1} {\vec{D}}^{1/2}_{\bgamma^{b}} = {\vec{Z}}\left({\vec{Z}}^{\rm H} {\vec{Z}}\right)^{-1}.
	\end{aligned}
	\end{equation}
	Note that this ZF matrix will suppress interference not only between CUs, as in conventional massive MIMO, but also from D2D transmitters. After applying the ZF detection matrix to the received signal~\eqref{eq:data BS-cu matrix}, we have a lower bound on the capacity of the CU~$k$ in cell $b$ as
	\begin{equation}
	\begin{aligned}\label{eq:achivebale ZF dedicated}
		R^{b,\rm c}_{b,k} & =  \left(1 - \frac{\tau }{\tau_{c}}\right)\log_2 \left(1+\frac{p^{\rm c}_{b,k} \gamma^{b,\rm c}_{b,k} (M-(K+N)) }{I^{\rm zf}_{b,k}} \right),\\
	\end{aligned}
	\end{equation}
	where
	\begin{equation}
	\begin{aligned}
		& I^{\rm zf}_{b,k} =1+ \sum\limits_{b'=1}^{B}\sum\limits_{k'=1}^{K} p^{\rm c}_{b',k'} \left( \beta^{b,\rm c}_{b',k'} -\gamma^{b,\rm c}_{b',k'}\right)\\
		&+ \sum\limits_{l=1}^{L} p^{b,\rm d}_{l} (\beta^{b,\rm d}_l -\gamma^{b,\rm d}_l)	+(M-(K+N)) \! \sum\limits_{b'=1, b' \neq b} p^{\rm c}_{b',k} \gamma^{b,\rm c}_{b',k}. 
	\end{aligned}
	\end{equation}
	Similar to the case of MR processing, here $\frac{p^{\rm c}_{b,k} \gamma^{b,\rm c}_{b,k} (M-(K+N)) }{I^{\rm zf}_{b,k}}$ is the effective SINR expression in which the numerator is the coherent beamforming gain of the desired signal at the BS $b$. In $I^{\rm zf}_{b,k}$, we have the noise variance, the non-coherent interference from CUs in contaminating cells, the non-coherent interference from D2D communication and  the coherent interference from CUs in other cell that are using the same pilot sequence as user $k$. Note that the details of the derivation will be present in the journal version of this paper.
	\subsection{Spectral efficiency of D2D communication}
	To calculate the SE of the D2D transmissions, we use the MMSE estimates provided in \eqref{eq:mmse dd-cu} and \eqref{eq:mmse dd}. To detect the desired data from D2D transmitter $l$, D2D receiver $l$ multiplies the received signal in \eqref{eq:data BS-dd} with the complex conjugate of its channel estimate $\hat{g}^{l,\rm d}_{l}$, which corresponds to MR processing. In addition, D2D receiver $l$ uses the channel estimates $\hat{g}^{l,\rm c}_{b,k}~\forall b, k$ and $\hat{g}^{l',\rm d}_{l}~\forall l'$ as side-information during data detection. Hence, by utilizing the lower bound for a fading channel with additive non-Gaussian noise and side information provided in \cite[Sec.~2.3]{redbook}, a lower bound on the capacity of the D2D communication of pair $l$ is obtained and presented as $R^{\rm d}_{l}$ in \eqref{eq:app bound dd dedicated} at the top of the next page. This expression is useful for SE computation but it is not in closed-form as it has an expectation with respect to the small-scale fading coefficients, thus it is not tractable for power control optimization. Hence, to provide a tractable power control algorithm, we compute an approximation of \eqref{eq:app bound dd dedicated} by computing the expectation of the numerator and the denominator of the fraction inside the logarithm. The resulting approximation of the SE of D2D pair $l$ is denoted as $\tilde{R}^{\rm d}_{l}$ and is given in~\eqref{eq:app bound dd dedicated final} at the top of the next page. One can show that the approximation error is negligible for SEs' less than $3\,$[b/s/Hz], which is the relevant interval for max-min SE optimization.
	\begin{figure*}
		{{\be
				\begin{aligned}\label{eq:app bound dd dedicated}
					&R^{\rm d}_{l} =\left(1 - \frac{\tau }{\tau_{c}}\right) \times \\
					&{\mathbb E}\left\{ \log_2 \left(1+\frac{p^{\rm d}_l \left|\hat{g}^{l,\rm d}_{l}\right|^2}{ p^{\rm d}_l \left({\beta}^{l,\rm d}_{l} - {\gamma}^{l,\rm d}_{l} \right) +  \sum\limits_{b=1}^{B}\sum\limits_{k=1}^{K} p^{\rm c}_{b,k} \left(\left|\hat{g}^{l,\rm c}_{b,k}\right|^2 + \left({\beta}^{l,\rm c}_{b,k} - {\gamma}^{l,\rm c}_{b,k}\right)\right) + \sum\limits_{l'=1,l'\neq l }^{L} p^{\rm d}_{l'} \left(\left|\hat{g}^{l,\rm d}_{l'}\right|^2 + \left({\beta}^{l,\rm d}_{l'} - {\gamma}^{l,\rm d}_{l'}\right)\right)+1 }\right)\right\}\\
				\end{aligned}
				\ee
		}} \hrulefill
	\end{figure*}
	\begin{figure*} \be
		\begin{aligned}\label{eq:app bound dd dedicated final}
			R^{\rm d}_{l}& \approx \tilde{R}^{\rm d}_{l} = \left(1 - \frac{\tau }{\tau_{c}}\right) \log_2 \left(1+\frac{ p^{\rm d}_l {\gamma}^{l,\rm d}_{l}}{p^{\rm d}_l \left({\beta}^{l,\rm d}_{l} - {\gamma}^{l,\rm d}_{l} \right) +  \sum\limits_{b=1}^{B}\sum\limits_{k=1}^{K} p^{\rm c}_{b,k} {\beta}^{l,\rm c}_{b,k} + \sum\limits_{l'=1,l'\neq l }^{L} p^{\rm d}_{l'} {\beta}^{l,\rm d}_{l'} +1 }\right).
		\end{aligned}
		\ee
		\hrulefill
	\end{figure*}
	
	\section{Optimization of Power Allocation}
	\label{optimization}
	In this paper, the optimization objective is to maximize the minimum SE of the CUs in all cells as well as all the D2D pairs. The max-min problem is written in epigraph form as
	\begin{maxi!}[2]
		{\substack{\{p^{\rm d}_l,p^{\rm c}_{b,k}\},\lambda}}{\lambda \label{eq:opti-objective}}
		{\label{eq:opti-original-epigraph}}{}
		\addConstraint{{R^{b,\rm c}_k}\geq \lambda ~~\forall b,k,\quad{\tilde{R}^{\rm d}_l} \geq \lambda~~\forall l \label{eq:C1opti}}{}{}
		\addConstraint{ 0\leq p^{\rm d}_l}{\leq P_{\rm max}~~\forall l \label{eq:C2opti}}
		\addConstraint{ 0\leq p^{\rm c}_{b,k}}{\leq P_{\rm max}~~ \forall b,k, \label{eq:C3opti}}
	\end{maxi!}
	where $\lambda$ indicates the quality of service (QoS) that is guaranteed to all CUs and all D2D pairs and this variable is to be maximized. In addition, $P_{\rm max}>0$ is the maximum transmit power of the users. Note that we use the SE approximation in \eqref{eq:app bound dd dedicated final} for the D2D pairs to achieve a tractable problem formulation. To solve \eqref{eq:opti-original-epigraph}, we fix $\lambda$ and solve the resulting linear feasibility optimization problem
	\begin{maxi!}[2]
		{\substack{\{p^{\rm d}_l,p^{\rm c}_{b,k}\}}}{0 \label{eq:opti-objective2}}
		{\label{eq:opti-original-epigraph2}}{}
		\addConstraint{{R^{b,\rm c}_k}\geq \lambda ~~\forall b,k,\quad{\tilde{R}^{\rm d}_l} \geq \lambda~~\forall l \label{eq:C1opti1}}{}{}
		\addConstraint{ 0\leq p^{\rm d}_l}{\leq P_{\rm max},~~\forall l \label{eq:C2opti2}}
		\addConstraint{ 0\leq p^{\rm c}_{b,k}}{\leq P_{\rm max},~~ \forall b,k. \label{eq:C3opti2}}
	\end{maxi!}
	We can then perform a line search for $\lambda$ over the interval $ [0,\lambda^{\rm u}]$, where $\lambda^{\rm u}$ is the utopia point, such that the problem is infeasible at this point. It is given as 
	\begin{equation}
			\begin{split}
		\begin{aligned}	
			\lambda^{\rm u}=&	
			\underset{l, b, k}\min\Big\{\log_2\left(1+P_{\max}p^{\rm d}_l {\gamma}^{\rm d}_{l,l}\right),\\ &{\log_2\left(1+P_{\max}(M)\gamma^{b,\rm c}_{b,k}\right)}\Big\}\\
		\end{aligned}
		\end{split}
	\end{equation}
	in the case of MR processing and with ZF processing it is
	\begin{equation}
	\begin{split}
		\begin{aligned}	
			\lambda^{\rm u}=& \underset{l, b,k}\min \Big\{\log_2\left(1+P_{\max}p^{\rm d}_l {\gamma}^{\rm d}_{l,l}\right),\\
			&{\log_2\left(1+P_{\max}(M-({K+N}))\gamma^{b,\rm c}_{b,k}\right)}\Big\}
		\end{aligned}
		\end{split}
	\end{equation}
	We can apply the well-known bisection line search algorithm to solve the problem~\eqref{eq:opti-original-epigraph} as a sequence of linear feasibility problem of the type in \eqref{eq:opti-original-epigraph2}.
	
	\section{Numerical Results}
	In this section, we provide a numerical evaluation of the performance of the power control algorithm proposed in Section~\ref{optimization}. Note that the exact expressions of SEs are used to compute the SEs. The simulation setup consists of a multi-cell network with $9$ cells as illustrated in Fig.~\ref{fig:model}. In addition, we use a wrap-around technique to avoid edge effects. We assume that the BSs are located in a $1$\,km$^2$ area. Each BS serves two CUs which are randomly distributed with uniform distribution in the BS's coverage area. The network contains $10$ D2D pairs randomly distributed in the coverage area with uniform distribution. Since the pairs are distributed randomly, some cells will be more affected by D2D interference than others. We further assume that the distance between the transmitter and receiver of each D2D pair is $10$ meters. We use the three-slope path-loss model from \cite{AoTang} to model the large-scale fading in the network and the parameters of this model is chosen from \cite{ngo2017cell}. This model contains two reference distance $d_0$ and $d_1$, that are chosen to be $10\,$ and $50\,$ meters, respectively.
	The carrier frequency is $2\,$GHz, the bandwidth is $20\,$MHz and the coherence interval contains $200$ symbols.  The total number of pilots assigned for D2D pairs is $5$,  where each D2D pair picks one of them randomly. In addition, the noise variance is set to~$-94\,$dBm and the maximum transmit power of the CUs and D2D transmitters are chosen to be~$200\,$mW for both pilot and data transmission. All users transmit their pilot with maximum power and we perform power control for data transmission. The performance of the proposed power control scheme is compared with the case without D2D pairs for ZF processing at all BSs. In addition, we provide result for the case of maximum power transmission at both D2D transmitters and CUs. 
	\begin{figure}[t]
		\centering
		\includegraphics[width=1\columnwidth]{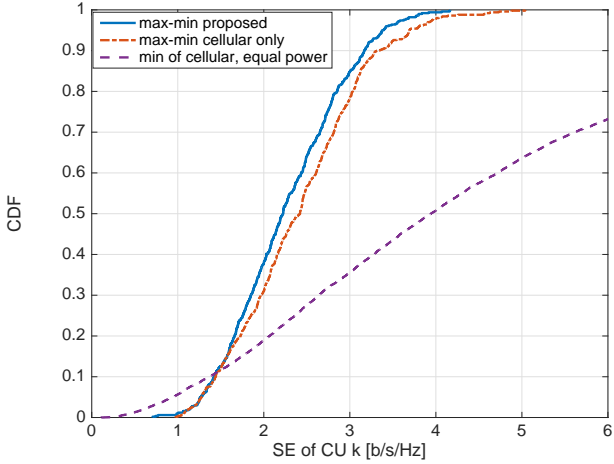} \vspace{-8mm}
		\caption{SE of arbitrary CU $k$} 
		\label{fig:SE CU}
	\end{figure} 
	Fig.~\ref{fig:SE CU} shows simulation results in terms of the cumulative distribution function (CDF) of the SE of a typical CU $k$. It can be seen that the performance of our proposed algorithm is almost the same as in the case  without D2D underlaying, thus the power control can efficiently mitigate the D2D interference. Comparing these results with the case of maximum power transmission, max-min power control improves the performance of the $10\%$ weakest users which is the goal of max-min fairness power control. Note that some fortunate users can get very high SE with maximum power transmission, but we did not show the results for values above 6\,b/s/Hz, since current standards do not support higher SEs than that.
	\begin{figure}[t]
		\centering
		\includegraphics[width=1\columnwidth]{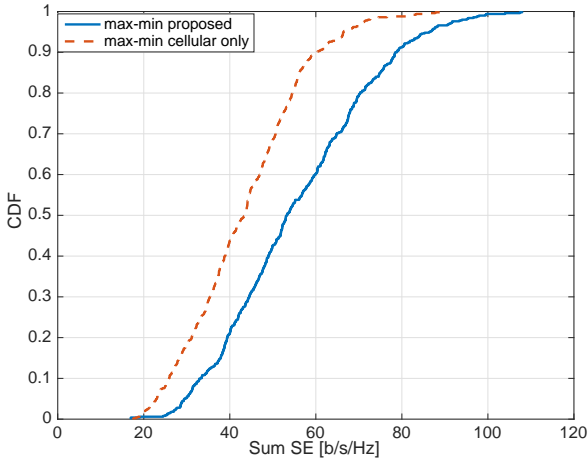} 
		\vspace{-8mm}
		\caption{Sum SE of network} \vspace{+2mm}
		\label{fig:Sum SE}
	\end{figure} 
	The gain of our proposed power control for D2D underlay is further shown in Fig.~\ref{fig:Sum SE}, where we compare the sum SE of the network for the D2D underlay case with the case of only cellular communication without D2D underlay. As we can see in this figure, the sum SE is higher with D2D underlaying. This is due to the fact that we have more active users in the network. We can conclude that D2D underlaying  multi-cell massive MIMO enhances the SE compared to conventional massive MIMO systems.

	\bibliographystyle{IEEEtran}
	
	\section{Conclusion}
	
	In this paper, we have presented a framework for pilot allocation in multi-cell Massive MIMO systems with D2D underlay. To enable interference mitigation at the cellular BSs, different orthogonal sets of pilots were used for cellular and D2D communication. We have shown how to estimate the channels in this setup and computed SE expressions for all users in the system using either MR or ZF processing. The proposed power control algorithm for max-min fairness optimization can effectively limit the interference between CUs and D2D pairs to enable D2D underlaying with limited performance degradation for the CUs.

	\bibliography{D2D_MIMO_multicell_WSA}

\end{document}

%% file: commands.tex
\newcommand{\be}{\begin{equation}}
\newcommand{\ee}{\end{equation}}
\newcommand{\bal}{\begin{aligned}}
\newcommand{\eal}{\end{aligned}}	
\newcommand{\ba}{\begin{array}}
\newcommand{\ea}{\end{array}}
\newcommand{\bea}{\begin{eqnarray}}
\newcommand{\eea}{\end{eqnarray}}

\newcommand{\vbar}{\raisebox{.17ex}{\rule{.04em}{1.35ex}}}
\newcommand{\vbarind}{\raisebox{.01ex}{\rule{.04em}{1.1ex}}}
\newcommand{\R}{\ifmmode {\rm I}\hspace{-.2em}{\rm R} \else ${\rm I}\hspace{-.2em}{\rm R}$ \fi}
\newcommand{\T}{\ifmmode {\rm I}\hspace{-.2em}{\rm T} \else ${\rm I}\hspace{-.2em}{\rm T}$ \fi}
\newcommand{\N}{\ifmmode {\rm I}\hspace{-.2em}{\rm N} \else \mbox{${\rm I}\hspace{-.2em}{\rm N}$} \fi}
\newcommand{\B}{\ifmmode {\rm I}\hspace{-.2em}{\rm B} \else \mbox{${\rm I}\hspace{-.2em}{\rm B}$} \fi}
\newcommand{\Hil}{\ifmmode {\rm I}\hspace{-.2em}{\rm H} \else \mbox{${\rm I}\hspace{-.2em}{\rm H}$} \fi}
\newcommand{\C}{\ifmmode \hspace{.2em}\vbar\hspace{-.31em}{\rm C} \else \mbox{$\hspace{.2em}\vbar\hspace{-.31em}{\rm C}$} \fi}
\newcommand{\Cind}{\ifmmode \hspace{.2em}\vbarind\hspace{-.25em}{\rm C} \else \mbox{$\hspace{.2em}\vbarind\hspace{-.25em}{\rm C}$} \fi}
\newcommand{\Q}{\ifmmode \hspace{.2em}\vbar\hspace{-.31em}{\rm Q} \else \mbox{$\hspace{.2em}\vbar\hspace{-.31em}{\rm Q}$} \fi}
\newcommand{\Z}{\ifmmode {\rm Z}\hspace{-.28em}{\rm Z} \else ${\rm Z}\hspace{-.28em}{\rm Z}$ \fi}

\renewcommand{\vec}[1]{\bf{#1}}     

\newcommand{\bgamma}{\mbox {\boldmath $\gamma$}}

\newcommand{\CN}{\mathcal{CN}}